\documentclass[twocolumn,showkeys,showpacs,prl]{revtex4}
\usepackage{amsfonts, amsmath, latexsym}
\usepackage[dvips]{graphicx}

\newtheorem{1}{Theorem}

\begin{document}

\title{The fast sampling algorithm for Lie-Trotter products}

\author{Cristian Predescu} 
\email{cpredescu@comcast.net}

\affiliation{
Department of Chemistry and Kenneth S. Pitzer Center for Theoretical Chemistry, University of California, Berkeley, California 94720
}
\date{\today}
\begin{abstract}
A fast algorithm for path sampling in path integral Monte Carlo simulations is proposed. The algorithm utilizes the L\'evy-Ciesielski implementation of Lie-Trotter products to achieve a mathematically proven  computational cost of $n\log_{2}(n)$ with the number of time slices $n$, despite the fact that each path variable is  updated separately, for reasons of optimality. In this respect, we demonstrate that updating a group of random variables simultaneously results in loss of efficiency. 
\end{abstract}
\maketitle

Path sampling in path integral Monte Carlo simulations  becomes more difficult in the limit of a large number of path variables or time slices, not only because of the build-up of correlation among the variables that are sampled, but also because of the shear increase in the number of random variables. Over the time, various approaches have been attempted in order to overcome the slowing down of the simulation. For direct sampling of Lie-Trotter products, among  the most successful are the staging method \cite{Spr85}, the threading algorithm \cite{Pol84}, the bisection method \cite{Cep95}, and the multigrid technique \cite{Jan93}. The so-called ``normal mode'' and Fourier formulations of path integrals have also been shown  to  improve sampling \cite{Dol84, Coa86, Fre86}. The last techniques are part of a larger class of path integral methods, class that is called the random series implementation \cite{Pre02}. To give a few examples, the computational effort in most random series approaches scales as $n^2$ with the number of path variables, whereas the bisection method  may reduce the effort down to $n^{1.4}$ \cite{Cep95}.

The random series approach, at least in the primitive form, is not  particularly efficient \cite{Pre02}. However, it reveals the incredibly large variety of possible different path integral formulations, while also being suggestive of more optimal approaches \cite{Pre04}. At the same time, it shows the strong connection that exists between these different forms, connection that is realized by means of certain orthogonal transformations. For Lie-Trotter products, Predescu and Doll \cite{Pre02b} have shown that there is an infinity of possible normal mode representations, which can be obtained one from the other by certain orthogonal transformations. One such transformation leads to the  L\'evy-Ciesielski representation, which has very special properties when it comes to numerical implementation. For instance, it allows for fast computation of paths, with a scaling of $n\log_{2}(n)$ for an entire path, as opposed to $n^2$, for other representations. The only restriction is that the number of time slices must be of the form $n = 2^k$, a condition reminiscent of the fast Fourier transform.
The algorithm we propose for the sampling of Lie-Trotter products utilizes the L\'evy-Ciesielski representation to achieve a similar goal: sampling in $n\log_{2}(n)$ operations of entire paths, while updating each path variable individually. Since any algorithm that computes entire paths every time has a computational effort proportional to $n$, we see that the proposed algorithm is highly efficient. Since it is not necessary to update all path variables all the time, the proposed algorithm can be made even more efficient by using it in conjunction with the multilevel Monte Carlo sampling technique \cite{Cep95}.

We commence by casting an arbitrary Lie-Trotter product in the L\'evy-Ciesielski form. In order to do so, we first enumerate the basic properties of the L\'evy-Ciesielski series representation of the Brownian bridge. For more details, the reader is advised to consult Refs.~\onlinecite{Pre04t,Pre02b, McK69}.   For $k=1,2,\ldots$ and $j=1,2,\ldots,2^{k-1}$, the Schauder functions $F_{k,j}(u)$ are generated by translations and dilatations of the function 
\begin{equation}
\label{eq:le1}
F_{1,1}(u) = \left\{\begin{array}{cc} u,& u \in (0, 1/2],\\ 1-u,& u \in (1/2, 1),\\ 0, &\text{elsewhere}. \end{array}\right.
\end{equation}
More precisely, we have
\begin{equation}
\label{eq:le2}
F_{k,j}(u)= 2^{-(k-1)/2} F_{1,1}(2^{k-1}u - j + 1),
\end{equation}
for  $k \geq 1$ and $1 \leq j \leq 2^{k-1}$.

\begin{figure}[!tbp] 
   \includegraphics[angle=270,width=8.5cm,clip=t]{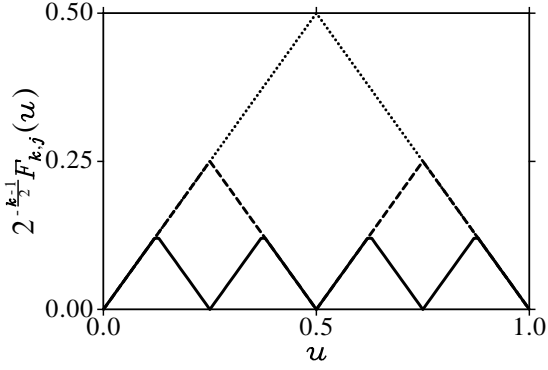} 
 \caption[sqr]
{\label{Fig:2}
A plot of the renormalized Schauder functions for the layers $k=1,2,\,\text{and}\,3$, showing the pyramidal structure.
}
\end{figure}

If we multiply them by $2^{-(k-1)/2}$, the Schauder functions make up a pyramidal structure organized in layers indexed by $k$, as shown in Fig.~\ref{Fig:2}.
The supports (the sets on which the functions do not vanish) of the Schauder functions are the open intervals of the form $(u_{k,j-1}, u_{k,j})$, for $1 \leq j \leq 2^{k-1}$, where $u_{k,j} = j2^{-(k-1)}$. The supports are \emph{disjoint} for functions corresponding to the same layer $k$. Because of this property, we have the equality 
\begin{equation}
\label{eq:le3}
\sum_{j = 1}^{2^{k-1}}a_{k,j} F_{k,j}(u) = a_{k, [2^{k-1}u]+1}F_{k,[2^{k-1}u] + 1}(u),
\end{equation}
for any sequence of numbers $a_{k,1}, a_{k,2}, \ldots , a_{k,2^{k-1}}$. Here, $[x]$ denotes the largest integer smaller or equal to $x$, whereas for $u = 1$, the quantities $a_{k, 2^{k-1} + 1}$ and $F_{k, 2^{k-1}+1}(1)$ are defined to be equal to $0$.

Let $\{a_{k,j}; k=1,2,\ldots; j=1,2,\ldots,2^{k-1}\}$ be an infinite sequence of independent identically distributed standard normal variables. From  Eq.~(\ref{eq:le3}) and the L\'evy-Ciesielski construction of the Brownian bridge \cite{Pre02b, Pre04t,McK69}, we have that 
\begin{equation}
\label{eq:le4} 
B^0_u\stackrel{d}{=} \sum_{k=1}^{\infty}a_{k, [2^{k-1}u]+1} F_{k,[2^{k-1}u] + 1}(u).
\end{equation}
In words, the right-hand side random series is equal in distribution to a standard Brownian bridge.

Let $n = 2^k -1$ be a fixed number and consider the equidistant points $u_{j} = j2^{-k}$, with  $1 \leq j \leq 2^k - 1 = n$ (these points were denoted before by $u_{k+1,j}$, but we shall drop the index $k + 1$ to avoid cluttering the formulas). Because the Schauder functions $F_{l,i}(u)$ vanish at these points for all levels $l \geq k + 1$, it follows that the  random sums
\[
 \sum_{l=1}^{k}a_{l, [2^{l-1}u_{j}]+1}F_{l,[2^{l-1}u_{j}] + 1}(u_{j})
\] 
for $j = 1, 2, \ldots, n$ have joint distribution equal to the joint distribution of  the random variables $B^0_{u_{j}}$. By the definition of the Brownian bridge, the joint distribution of the latter variables is given by the formula
\begin{equation}
\label{eq:le7}
\frac{1}{p_1(0,0)} p_{u_{1}}(0, x_1)p_{u_{2} - u_{1}}(x_1, x_2) \cdots p_{1 - u_{n}}(x_{n}, 0),
\end{equation}
with $p_u(x,x')$ defined by 
\begin{equation}
\label{eq:le6}
p_u(x,x') = (2\pi u)^{-1/2}\exp\left[-{(x'-x)^2}/({2u})\right].
\end{equation}
Using this observation, the notation $x_r(u) = x + (x'-x)u$, and the definition $\sigma^2 = \hbar^2 \beta /m_0$, one easily proves that the joint distribution of the random variables
\begin{equation}
\label{eq:le8}
x_r(u_j) + \sigma  \sum_{l=1}^{k}a_{l, [2^{l-1}u_{j}]+1}F_{l,[2^{l-1}u_{j}] + 1}(u_{j})
\end{equation}
for $j = 1, 2, \ldots, n$ is given be the formula
\begin{eqnarray}
\label{eq:le9} \nonumber &&
p_{\sigma^2 u_{1}}(x, x_1)p_{\sigma^2(u_{2} - u_{1})}(x_1, x_2) \\ && \cdots p_{\sigma^2(1 - u_{n})}(x_{n}, x') \left/ p_{\sigma^2}(x,x') \right. .
\end{eqnarray}

Because the density matrix of a free particle is strictly positive, any short-time approximation $\rho_0(x,x';\beta)$ can be put in the product form
\begin{equation}
\label{eq:le10}
\rho_0(x,x';\beta) = \rho_{fp}(x,x';\beta)r_0(x,x';\beta). 
\end{equation}
Letting $x_0 = x$, $x_{n+1} = x'$, $u_0 = 0$, and $u_{n+1} = 1$,  the $n$-th order  Lie-Trotter product obtained from the short-time approximation considered above takes the form
\begin{eqnarray}
\label{eq:le11}
\nonumber
\rho_n(x,x';\beta) =\int_{\mathbb{R}^n}  \prod_{i = 0}^{n}p_{\sigma^2(u_i - u_{i+1})}(x_i,x_{i+1}) \\ \times \prod_{j = 0}^n r_0(x_{j},x_{j+1};\beta/2^k) dx_1 \cdots dx_n. 
\end{eqnarray}
From the last equation and the fact that the distribution given by Eq.~(\ref{eq:le9}) is the distribution of the random variables appearing in Eq.~(\ref{eq:le8}), we readily obtain the following \emph{L\'evy-Ciesielski form of Lie-Trotter products} for $n + 1 = 2^k$ time slices
\begin{widetext}
\begin{eqnarray}
\nonumber 
\label{eq:le12}
\rho_n(x,x';\beta) &=&\rho_{fp}(x,x';\beta) \int_{\mathbb{R}}da_{1,1} \cdots \int_{\mathbb{R}}da_{k,2^{k-1}} (2\pi)^{-n/2}\prod_{l = 1}^{k}\prod_{i = 1}^{2^{l-1}}\exp\left(-a_{l,i}^2/2\right)  \\ &&\times \prod_{j = 0}^{n} r_0\left[x_r(u_j) + \sigma \sum_{l=1}^{k}a_{l, [2^{l-1}u_{j}]+1}F_{l,[2^{l-1}u_{j}] + 1}(u_{j}), \right.\\&& \left.  x_r(u_{j+1}) + \sigma \sum_{l=1}^{k}a_{l, [2^{l-1}u_{j+1}]+1}F_{l,[2^{l-1}u_{j+1}] + 1}(u_{j+1});\beta/2^k\right]. \nonumber
\end{eqnarray}
\end{widetext}
This formula has been first introduced in Ref.~\cite{Pre02b}, albeit for some specialized short-time approximations.

We have already mentioned that one of the advantages of the L\'evy-Ciesielski form for Lie-Trotter products is that it enables \emph{fast computation of paths} \cite{Pre02b}, while maintaining the random series appearance of the final expression. This is so because for each $u_j$, one needs to perform a number of $k = \log_2(n+1)$ operations in order to compute the coordinate $u_j$ of the path. This translates into a scaling of $(n+1)\log_2(n+1)$ operations for a whole path.  

As announced in the beginning of the letter, a second advantage of the L\'evy-Ciesielski form is that it allows for \emph{fast Monte Carlo sampling} of the distribution given by the right-hand side of Eq.~(\ref{eq:le12}). The  algorithm is as follows. A trial move is proposed for all $2^{l-1}$ variables $a_{l,i}$ that correspond to a single level $l$. However, the acceptance/rejection decision is taken individually for each path variable, because these variables are statistically independent. To see this, first observe that the part of the product distribution in Eq.~(\ref{eq:le12}) that only involves the path variables for the layer $l$ factorizes as 
\begin{widetext}
\begin{eqnarray}
\nonumber 
\label{eq:fa1}
\prod_{i = 1}^{2^{l-1}} \left\{(2\pi)^{-1/2}e^{-a_{l,i}^2/2}  \prod_{j = (i-1)2^{k-l+1} }^{i2^{k-l+1} - 1 } r_0\left[x_r(u_j) + \sigma \sum_{l=1}^{k}a_{l, [2^{l-1}u_{j}]+1}F_{l,[2^{l-1}u_{j}] + 1}(u_{j}), \right. \right.\\ \left. \left. x_r(u_{j+1}) + \sigma \sum_{l=1}^{k}a_{l, [2^{l-1}u_{j+1}]+1}F_{l,[2^{l-1}u_{j+1}] + 1}(u_{j+1});\beta/2^k\right]\right\}. 
\end{eqnarray}
\end{widetext}
No two factors defined by the curly brackets contain a same variable $a_{l,i}$. Indeed, given $i \in \{1,2,\ldots 2^{l-1}\}$, only the function $F_{l,i}(u)$ may be non-zero for the values $u_j$ with $(i-1)2^{k-l + 1}\leq j \leq i2^{k-l + 1}$. Thus, each factor defined by the curly brackets appearing in Eq.~(\ref{eq:fa1}) contains one and only one variable $a_{l,i}$. Therefore, the variables $a_{l,i}$ are and should be treated as \emph{independent} during the Monte Carlo simulation. Each proposal $a_{l,i} \to a'_{l,i}$ \emph{must be tested separately} using the weight given by the appropriate factor and accepted or rejected according to the  Metropolis-Hastings rule. 

Let us analyze the efficiency of the algorithm. First, there are $k = \log_2(n+1)$ layers. For each layer, one evaluates the function $r_0(x,x';\beta)$ exactly $n + 1$ times, in order to update all variables from the layer, individually.  Thus, the computational effort to update all variables individually is proportional to $(n+1)\log_2(n+1)$. Since any algorithm must have a scaling of  at least $n+1$ [this is the computational effort necessary to evaluate the distribution given by Eq.~(\ref{eq:le12}) for any update attempt], we see that the technique is extremely efficient, especially given that all variables are moved separately.  Of the variables $a_{l,i}$, the most difficult to sample is $a_{1,1}$, because its  distribution stretches over a larger region [the distance covered by a variable $a_{l,j}$ is of the order $\sigma 2^{-(l-1)/2}$, for low temperatures]. Therefore, the overall computational effort to perform an accurate sampling of all variables also scales as $(n+1)\log_2(n+1)$.

The reader may ask why  we have insisted on updating each path variable individually. The answer is that there is a loss of efficiency if we try to update more than one path variable at a time. We shall  prove this assertion in the remainder of the letter. Let's assume we are given a finite collection $X_1, X_2, \ldots, X_n$ of independent identically distributed random vectors (i.i.d.r.v's), taking values in some space $\mathbb{R}^d$.  Let $\rho(\mathbf{x})$, with $\mathbf{x} \in \mathbb{R}^d$, be the normalized distribution of any of the random vectors $X_i$. By independence, the overall distribution is given by the product $\rho(\mathbf{x}_1)\rho(\mathbf{x}_2)\ldots \rho(\mathbf{x}_n)$, which is a distribution on the space $\mathbb{R}^{dn}$. Assume we attempt to update all variables at once, using a trial distribution $T(\mathbf{y}_1|\mathbf{x}_1)T(\mathbf{y}_2|\mathbf{x}_2)\ldots T(\mathbf{y}_n|\mathbf{x}_n)$.
The move to $(\mathbf{y}_1, \mathbf{y}_2, \ldots, \mathbf{y}_n)$ is then accepted  with probability
\begin{equation}
\label{eq:co1}
\min\left\{1, \prod_{i = 1}^n \frac{\rho(\mathbf{y}_i)T(\mathbf{x}_i | \mathbf{y}_i)}{\rho(\mathbf{x}_i)T(\mathbf{y}_i | \mathbf{x}_i)}\right\},
\end{equation}
and rejected with the remaining probability. The average acceptance probability is  given by the formula
\begin{eqnarray}
\label{eq:co2} \nonumber
\mathrm{Ac}(n) = \int_{\mathbb{R}^{2d}}d\mathbf{x}_1d\mathbf{y}_1 \cdots \int_{\mathbb{R}^{2d}}d\mathbf{x}_nd\mathbf{y}_n \rho(\mathbf{x}_1)T(\mathbf{y}_1|\mathbf{x}_1)\\ \cdots \rho(\mathbf{x}_n)T(\mathbf{y}_n|\mathbf{x}_n)
\min\left\{1, \prod_{i = 1}^n \frac{\rho(\mathbf{y}_i)T(\mathbf{x}_i | \mathbf{y}_i)}{\rho(\mathbf{x}_i)T(\mathbf{y}_i | \mathbf{x}_i)}\right\}. 
\end{eqnarray}

In these conditions, we have the following theorem that guarantees that simultaneous sampling of i.i.d.r.v's  is inefficient.
\begin{1}[Bad sampling of i.i.d.r.v's]
\label{th:co1}
Except for the ideal case $T(\mathbf{y}|\mathbf{x}) = \rho(\mathbf{y})$ whenever $T(\mathbf{y}|\mathbf{x}) \neq 0$, there is a strictly positive constant $H$ such that 
\begin{equation}
\label{eq:co3}
\mathrm{Ac}(n)\sim e^{-Hn}. 
\end{equation}
This constant is given by the relative Shannon entropy
\begin{equation}
\label{eq:co4}
H = - \int_{\mathbb{R}^{2d}}\rho(\mathbf{x})T(\mathbf{y}|\mathbf{x}) \log\left[\frac{\rho(\mathbf{y})T(\mathbf{x}| \mathbf{y})}{\rho(\mathbf{x})T(\mathbf{y} | \mathbf{x})}\right]d\mathbf{x}d\mathbf{y}.
\end{equation}
\end{1}

\emph{Proof of the theorem.} For convenience, we let $\mathbb{E}$ denote the expected  value against the distribution $\rho(\mathbf{x})T(\mathbf{y}|\mathbf{x})$, i.e., for some arbitrary function $f(\mathbf{x},\mathbf{y})$,
\[
\mathbb{E}\left[f(X,Y)\right] = \int_{\mathbb{R}^{2d}} \rho(\mathbf{x})T(\mathbf{y}|\mathbf{x})f(\mathbf{x},\mathbf{y})d\mathbf{x}d\mathbf{y}.
\] 
Then Eq.~(\ref{eq:co2}) can be written as
\begin{equation} 
\label{eq:co5}
\mathrm{Ac}(n) = \mathbb{E}_1\cdots\mathbb{E}_n 
\min\left\{1, \prod_{i = 1}^n \frac{\rho(Y_i)T(X_i | Y_i)}{\rho(X_i)T(Y_i | X_i)}\right\}. 
\end{equation}

Now, consider the identity
\begin{eqnarray}
\label{eq:co6}
&& \nonumber
\prod_{i = 1}^n \frac{\rho(Y_i)T(X_i | Y_i)}{\rho(X_i)T(Y_i | X_i)} \\ && = \exp\left(-n \left\{-\frac{1}{n}\sum_{i = 1}^n \log\left[\frac{\rho(Y_i)T(X_i | Y_i)}{\rho(X_i)T(Y_i | X_i)}\right]\right\}\right).
\end{eqnarray}
The expression inside the curly brackets is a ``time'' average of independent identically distributed random variables. By the law of large numbers, this time average will converge to a constant function, the value of which is the ``space'' average 
\begin{eqnarray}
\nonumber
\label{eq:co7}
H = - \lim_{n \to \infty}\frac{1}{n}\sum_{i = 1}^n \log\left[\frac{\rho(Y_i)T(X_i | Y_i)}{\rho(X_i)T(Y_i | X_i)}\right] \\ = - \mathbb{E}\left\{ \log\left[\frac{\rho(Y)T(X | Y)}{\rho(X)T(Y | X)}\right]\right\}.
\end{eqnarray}
Remembering the definition of $\mathbb{E}$, we see that the right-hand side of the previous equation is the Shannon entropy of the probability measure $\rho(\mathbf{x})T(\mathbf{y}|\mathbf{x})$  relative to the measure $\rho(\mathbf{y})T(\mathbf{x}|\mathbf{y})$. It is always non-negative and, in fact, it is (strictly) positive except for the case
\begin{equation}
\label{eq:co8}
\rho(\mathbf{y})T(\mathbf{x}|\mathbf{y}) = \rho(\mathbf{x})T(\mathbf{y}|\mathbf{x}).
\end{equation}

To prove the last assertions, notice that $-\log(x)$ is a strictly convex function and remember Jensen's inequality, which in this case says
\[
\mathbb{E}\left\{-\log[f(X,Y)]\right\} \geq - \log\left\{\mathbb{E}\left[f(X,Y)\right] \right\},
\]
with equality if and only if $f(X,Y)$ is constant. Then, 
\begin{eqnarray*} &&
H = - \int_{\mathbb{R}^{2d}}\rho(\mathbf{x})T(\mathbf{y}|\mathbf{x}) \log\left[\frac{\rho(\mathbf{y})T(\mathbf{x}| \mathbf{y})}{\rho(\mathbf{x})T(\mathbf{y} | \mathbf{x})}\right]d\mathbf{x}d\mathbf{y} \\ &&  \geq  - \log\left[ \int_{\mathbb{R}^{2d}}\rho(\mathbf{x})T(\mathbf{y}|\mathbf{x})\frac{\rho(\mathbf{y})T(\mathbf{x}| \mathbf{y})}{\rho(\mathbf{x})T(\mathbf{y} | \mathbf{x})} d\mathbf{x}d\mathbf{y}\right] = 0.
\end{eqnarray*}
Eq.~(\ref{eq:co3}) follows from Eqs.~(\ref{eq:co5}), (\ref{eq:co6}), and (\ref{eq:co7}), together with the observation that $\min\{1, \exp(-Hn)\} = \exp(-Hn)$, since $H \geq 0$. 
Eq.~(\ref{eq:co8}) follows from the second part of Jensen's inequality. It is readily seen to be equivalent to the statement $\rho(\mathbf{y}) = T(\mathbf{y}|\mathbf{x})$, whenever $T(\mathbf{y}|\mathbf{x}) \neq 0$. The proof of the theorem is concluded. 

We have therefore demonstrated that there is a  loss of efficiency if simultaneous updating of path variables is attempted, even for uncorrelated variables. 
If simultaneous sampling is performed, one must modify the proposal $T(\mathbf{y}|\mathbf{x})$, so that  the Shannon entropy decreases with the dimensionality at a rate faster than $O(1/n)$. Straightforward calculations show that, if the proposal $T(\mathbf{y}|\mathbf{x})$ is uniform in  a $d$-dimensional hypercube centered about the previous position, the maximal displacements must be decreased at a rate faster than or equal to $O(n^{-1/2})$, in order to prevent the severe degradation of the quality of the simulation predicted by Th.~\ref{th:co1}.  For arbitrary random series, there is little choice: either we decrease the maximal displacements or update each path variable individually, at a total cost proportional to $n^2$. 

However, for the L\'evy-Ciesielski representation, the random variables can be updated individually in $n\log_2(n)$ operations, where $n$ is the number of time slices. For this reason, as well as for the property of fast computation of paths, we believe that the sampling algorithm we have presented will prove to be a valuable tool for all path integral simulations that are implemented via Lie-Trotter products. Again, the main property of the L\'evy-Ciesielski representation that has enabled the development of this algorithm is the fact that the path variables corresponding to a same layer are statistically independent.

\begin{acknowledgments} This work was supported in part by the National Science Foundation Grant Number CHE-0345280, the Director, Office of Science, Office of Basic Energy Sciences, Chemical Sciences, Geosciences, and Biosciences Division, U.S. Department of Energy under Contract Number DE AC03-65SF00098, and the U.S.-Israel Binational Science Foundation Award Number 2002170.
\end{acknowledgments}

\end{document}